\documentstyle[12pt]{article}

\def\gev{{\hbox{GeV}}}

\begin{document}
\baselineskip 7.2mm

\begin{titlepage}

\begin{flushright}
DPNU-93-16 \\
AUE-03-93 \\
July\ \ 1993 \\
\end{flushright}

\vspace {1cm}

\begin{center}
{\Large{\bf The String Unification of Gauge Couplings \\
and Gauge Kinetic Mixings }}

\vspace {1cm}

Chuichiro HATTORI \\
{\it Department of Physics, Aichi Institute of Technology \\
Toyota, Aichi, JAPAN 470-03}

\vspace {3mm}
Masahisa MATSUDA \\
{\it Department of Physics and Astronomy \\
Aichi University of Education \\
Kariya, Aichi, JAPAN 448}

\vspace {3mm}
Takeo MATSUOKA and Daizo MOCHINAGA \\
{\it Department of Physics, Nagoya University \\
Nagoya, JAPAN 464-01}
\end{center}

\vspace {1cm}

\begin{abstract}
In the superstring models we have not only the complete {\bf 27}
multiplets of $E_6$ but also extra incomplete
$({\bf 27}+{\overline {\bf 27}})$
chiral supermultiplets being alive at low energies.
Associated with these additional multiplets,
when the gauge symmetry
contains more than one $U(1)$ gauge group,
there may exist gauge kinetic mixings among these $U(1)$ gauge groups.
In such cases the effect of gauge kinetic mixings should
be incorporated into the study of unification of gauge couplings.
We study these interesting effects systematically
in these models.
The string threshold effect is also taken into account.
It is found that in the four-generation models we do not
have a advisable solution of string unification of gauge couplings
consistent with experimental values at the electroweak scale.
We also discuss the possible scenarios to solve this problem.
\end{abstract}

\end{titlepage}

\section{Introduction}
\hspace*{\parindent}
Recent precise data at LEP have called back renewed attention
to the unification of gauge coupling constants.
It was found that the unification of gauge couplings
$g_3$, $g_2$ and $g_1$
of $SU(3)\times SU(2)\times U(1)$ occurs
at about $10^{16}$GeV in the minimal supersymmetric
standard model (MSSM)
\cite{Amaldi}.
On the other hand, at present, the most promising unified
theory including gravity is the superstring theory.
In superstring derived models, however, the unification scheme
of gauge couplings is apparantly different from MSSM.
First, in superstring models gauge couplings are expected to be
unified at the string scale of about $10^{18}$GeV, rather than
about $10^{16}$GeV.
In addition, string threshold corrections play an important role
to interpret the unification at about $10^{18}$GeV.
Second, in superstring models the gauge symmetry at the unification
scale is rank-6 or rank-5 and is larger than the standard gauge
symmetry $G_{st}=SU(3)_C\times SU(2)_L\times U(1)_Y$ with rank-4.
Consequently there may exist intermediate energy scales of
symmetry breaking between the unification scale and
the electroweak scale.
Third, in the superstring models matter contents are not minimal.
Additional matter fields which do not exist in the minimal model
like the MSSM
are involved in {\bf 27}-representation of $E_6$.
Further, we have not only $N_f$(family number) sets of complete {\bf 27}
multiplet but also extra incomplete
$({\bf 27}+{\overline {\bf 27}})$ multiplets
as chiral superfields available at low energies.
When we investigate gauge unification in superstring models,
we have to take these non-standard features into account.

In Ref.%
\cite{Tsunoda}
the string unification of non-abelian gauge couplings has been studied
for Calabi-Yau models with one K\"ahler class modulus.
It was found that the threshold effects lead to an important unification
condition on gauge couplings.
This condition is that the energy scale at which
gauge couplings have a common value
should not be smaller than $\sim 10^{18}$GeV.
By analyzing the running of non-abelian gauge couplings in
four-generation superstring models,
it was shown that there is a consistent model in which
non-abelian gauge couplings join at about $10^{19}$GeV.
In this model the gauge symmetry
at the string scale is $SU(4)_C\times SU(2)_L\times U(1)\times U(1)$.
In four-generation models the intermediate scale $M_I$ of symmetry
breaking is naturally bounded to be about $10^{16}$GeV
\cite {4generation}.
At this scale $M_I$ the gauge symmetry is spontaneously broken
into $SU(3)_C\times SU(2)_L\times U(1)\times U(1)$.
As seen in this model, the superstring derived models mostly contain
more than one $U(1)$ gauge group at various stages
of symmetry breaking.
In such cases there may exist
gauge kinetic mixing terms in the effective Lagrangian
\cite{Choi}.
In general, we have massless and/or massive modes which are charged
states in two or more $U(1)$ gauge groups.
If we have only complete {\bf 27} multiplets, the gauge kinetic mixing term
does not appear as a consequence of the summation over the complete 27
fields.
However, extra incomplete $({\bf 27}+{\overline {\bf 27}})$
multiplets possibly
contribute to the gauge kinetic mixing at one-loop level.
In the presence of such mixing it is necessary for us to
diagonalize the gauge kinetic terms.
In this paper we focus our attention on the study of unification
of gauge couplings with gauge kinetic mixings.
In the study we also take the string threshold corrections
into account.
Concretely, we take up the four-generation Calabi-Yau models
and solve kinetic mixing problem of abelian gauge couplings.

This paper is organized as follows.
In section 2 we briefly discuss the string threshold corrections
and a unification condition for the Calabi-Yau models with
one K\"ahler class modulus.
It is pointed out that in the evolution of abelian gauge couplings
we should pay attention to gauge kinetic mixing,
which generally appears in the effective theories with more than
one $U(1)$ gauge group.
In section 2 we also study the diagonalization of the coefficients
of the $\beta $-function in the effective theories with
gauge kinetic mixings.
To make the present study concrete,
we take up the four-generation models in section 3.
In one of the four-generation models, in which the unification
condition is fulfilled for non-abelian gauge couplings,
gauge kinetic mixing takes place at the string scale
by the effect of additional incomplete $({\bf 27}+{\overline {\bf 27}})$
multiplets and also at the intermediate energy scale
as the effects that partial multiplets of complete {\bf 27}
multiplet become massive at the intermediate scale
and that additional incomplete $({\bf 27}+{\overline {\bf 27}})$
multiplets are still massless.
In section 4 we carry out the renormalization group analysis
for the above four-generation model.
The emphasis is placed on solving the
gauge kinetic mixing problem.
We explore advisable solutions which are consistent with
experimental values of gauge couplings at the electroweak scale.
Section 5 is devoted to summary and discussion.

\section{Threshold Corrections and Kinetic Mixings}
\hspace*{\parindent}
In the superstring theory we have the target space duality symmetry,
which interchanges Kaluza-Klein modes and
winding modes of the compactified string.
In the case of only one modulus field $T$ the duality symmetry
requires the invariance under the $PSL(2,{\bf Z})$ transformation
\cite{Kikkawa}
\begin{equation}
   T \longrightarrow \frac{aT-ib}{icT+d}
         \ \ \ \ \ \ \ \ a,b,c,d \in {\bf Z}
               \ \ \ \ \ ad-bc=1.
\end{equation}
In the Calabi-Yau models with one K\"ahler class modulus
the duality symmetry gets the string threshold correction
into a simple form.
As discussed in Ref.%
\cite {Tsunoda},
non-abelian gauge coupling is written down as
\begin{equation}
    \alpha _a^{-1}(\mu )=\alpha _{st}^{-1}+\frac{b_a}{4\pi }
       \left\{
          \ln \left(\frac{M_C^2}{\mu ^2}\right) -f(T,\overline {T})
       \right\}.
\end{equation}
at one-loop level, where $\alpha _{st}\equiv g_{st}^2/(4\pi )$ is
a universal constant independent of the various gauge groups $G_a$.
In this equation the string threshold correction is described
in terms of the K\"ahler class moduli-dependent function
$f(T,\overline{T})$.
$T$ is related to the size $R$ of the compactified manifold
as $\langle {\rm Re}\,T\rangle =2R^2$ in unit of $(\alpha ')^{1/2}$,
where $(2\pi \alpha ')^{-1}$ means the string tension.
In Eq.(2), $b_a$ stands for a coefficient of the one-loop
$\beta $-function given by
\begin{equation}
    b_a=-3C_2(G_a)+\sum _fT(R_f).
\end{equation}
In the $\overline {DR}$ scheme the string scale $M_C$ is determined as
\begin{equation}
    M_C=\left( \frac{2\exp(1-\gamma _E)}
    	{3\sqrt {3}\pi \alpha '}\right)^{1/2}
          =0.73\times g_{st}\times 10^{18}{\rm GeV},
\end{equation}
where $\gamma _E$ is the Euler constant.
According to the duality symmetry $f(T,\overline{T})$ is of the form
\cite{Kaplunovsky}
\begin{equation}
    f(T,\overline{T})=
        \ln \left\{ (T+\overline{T})\left| \eta (iT)\right|^4\right\}.
\end{equation}
Here $\eta (iT)$ is the Dedekind's $\eta $-function given by
\begin{equation}
    \eta (iT)=\exp\left(\frac {-\pi T}{12}\right)
          \prod _{n=1}^{\infty}\left( 1-\exp(-2\pi nT)\right).
\end{equation}
The moduli space of the $T$-field
can be taken as the so-called fundamental domain
$-\frac {1}{2}\leq {\rm Im}T < \frac{1}{2},\  |T|\geq 1$.
The term $\ln (T+\overline{T})$ in $f(T,\overline{T})$ represents
the contribution only from one-loop with massless modes.
While the term $\ln |\eta (iT)|^4$ comes from one-loop effects with
massive modes.
Although each term has a duality anomaly,
the anomaly cancels out with each other.
{}From properties of $\eta $-function we obtain
\begin{equation}
    f(T,\overline{T}) \approx -\frac{\pi }{3}{\rm Re}\,T
                              +\ln \left( 2{\rm Re}\,T\right)
                          -4\exp(-2\pi{\rm Re}\,T)\cos(2\pi{\rm Im}\,T)
                      \leq  -0.34
\end{equation}
in the fundamental domain of the moduli space.

Now we introduce the unphysical parameter $M_X$
as the energy scale at which
non-abelian gauge couplings have a common value.
Following this definition, we have
\begin{equation}
    \ln \left( \frac{M_C^2}{M_X^2} \right) =f(T,\overline{T})
\end{equation}
at one-loop level.
The constraint (7) on $f(T,\overline T)$ implies an inequality
\begin{equation}
    M_X = M_C\exp \left(-\frac{1}{2}f(T,\overline T)\right) >
    0.87 \times g_{st} \times 10^{18}{\rm GeV}.
\end{equation}
This is an important unification condition on Calabi-Yau
models with the anti-generation number $h^{11}=1$.

Next we proceed to discuss abelian gauge couplings.
If we have only one $U(1)$ gauge group, the renormalization
group analysis is completely parallel to those for non-abelian
gauge couplings.
In the superstring derived models, however, the gauge group
at the string scale is rank-6 or rank-5 and mostly contains
more than one $U(1)$ gauge group.

For illustration let us consider the case in which
the gauge group at the string scale
$M_C$ contains two $U(1)$ groups, i.e. $U(1)_A\times U(1)_B$.
At tree level gauge couplings $g_A$ and $g_B$ take
a universal value $g_{st}$.
Here we denote $U(1)_{A,B}$-charges of
massless fields $f$ as $q_f^{(A,B)}$,
which are normalized as
\begin{equation}
    \sum_{f\in {\bf 27}}\left(q_f^{(A)}\right)^2=
    \sum_{f\in {\bf 27}}\left(q_f^{(B)}\right)^2=3.
\end{equation}
Under this normalization, tree-level abelian gauge couplings have also
a common value together with tree-level non-abelian gauge couplings.
We introduce the notation
\begin{equation}
     b_{ij}=\sum_{f\in \Psi _0} q_f^{(i)} q_f^{(j)}
                        \ \ \ \ (i,j=A,B).
\end{equation}
The summation should be taken over all massless fields $\Psi _0$.
Generally, there exists the gauge kinetic mixing term
$F_{\mu \nu}^{(A)}F^{(B)\mu \nu }$ at one-loop level as shown in Fig.1.
The magnitude of this mixing term is proportional to $b_{AB}$.
If we confine the summation to one set of
massless particles constructing
complete {\bf 27} multiplet of $E_6$, we have
\begin{equation}
    b_{AB} = \sum _{f\in {\bf 27}}q_f^{(A)}q_f^{(B)} = 0.
\end{equation}
\bigskip
\begin{center}
         {\large {\bf Fig.1}}
\end{center}
\bigskip
Then, when we have only complete {\bf 27} multiplets,
the gauge kinetic mixing of two abelian groups does not
occur at one-loop level.
However, in the superstring models there appear not only
$N_f$(family number) sets of complete {\bf 27} multiplet but also
extra incomplete $({\bf 27}+{\overline {\bf 27}})$ chiral supermultiplets.
These additional $({\bf 27}+{\overline {\bf 27}})$ multiplets
possibly contain charged states both in $U(1)_A$ and $U(1)_B$
and generally $b_{AB}$ becomes nonvanishing, i.e.
\begin{equation}
   b_{AB}=\sum _{f\in \Psi _0}q_f^{(A)}q_f^{(B)}\neq 0.
\end{equation}

In order to diagonalize $b_{ij}$, we carry out the orthogonal
transformation $U(1)_A\times U(1)_B\rightarrow U(1)_D\times U(1)_E$, i.e.
\begin{eqnarray}
                                        \left(
   \begin{array}{l}
         D_{\mu } \\
         E_{\mu }
   \end{array}
                                        \right)
                                     & = &
                                        \left(
   \begin{array}{cc}
         \cos \omega & \sin \omega \\
        -\sin \omega & \cos \omega
   \end{array}
                                        \right)
                                        \left(
   \begin{array}{l}
         A_{\mu } \\
         B_{\mu}
   \end{array}
                                        \right), \nonumber \\
                                        \left(
   \begin{array}{l}
         Q_D \\
         Q_E
   \end{array}
                                        \right)
                                    & = &
                                        \left(
   \begin{array}{cc}
         \cos \omega & \sin \omega \\
        -\sin \omega & \cos \omega
   \end{array}
                                        \right)
                                        \left(
   \begin{array}{l}
         Q_A \\
         Q_B
   \end{array}
                                        \right),
\end{eqnarray}
where $A_{\mu }$, $B_{\mu }$, $D_{\mu }$ and $E_{\mu }$ represent
gauge fields and $Q_i$ $(i=A,B,D,E)$ are $U(1)_i$ generators.
When the rotation angle $\omega $ is taken as
\begin{equation}
         \tan 2\omega = \frac{2b_{AB}}{b_{AA}-b_{BB}}\
\label{eqn:omega}
\end{equation}
and $-\pi/4 \leq \omega \leq \pi/4$,
$b_{ij}$ is diagonalized as
\begin{eqnarray}
    b_{DE} &=& 0,                                      \nonumber \\
    b_{DD} &=& \frac{1}{2}(b_{AA}+b_{BB})
                +\frac{b_{AA}-b_{BB}}{2\cos 2\omega }, \\
    b_{EE} &=& \frac{1}{2}(b_{AA}+b_{BB})
                -\frac{b_{AA}-b_{BB}}{2\cos 2\omega }. \nonumber
\end{eqnarray}
In this basis we can easily solve the one-loop
renormalization group equations.
Combining the string threshold corrections, we have
\begin{equation}
    \alpha _i^{-1}(\mu )=\alpha _{st}^{-1}+\frac{b_{ii}}{4\pi }
        \ln \left(\frac{M_X^2}{\mu ^2}\right)
            \ \ \ \ \ \ (i=D,E)
\end{equation}
at one-loop level.

In many of the superstring derived models,
there possibly exists an intermediate scale
of symmetry breaking between the unification scale $M_C$ and
the electroweak scale $M_Z$.
In the presence of the symmetry breaking we are potentially
confronted with the gauge kinetic mixing problem.
Here let us consider the case in which the symmetry breaking
takes place at the scale $M_I$ and the gauge symmetry at energies
below $M_I$ contains two $U(1)$ gauge groups
$U(1)_G\times U(1)_H$.
Through the symmetry breaking some fields gain masses of order $M_I$
and the others remain massless.
These massive fields decouple from the effective theory below $M_I$.
Therefore, there is a possibility
that below $M_I$ we obtain
\begin{equation}
    b'_{GH}=\sum_{f\in \Psi '_0} q_f^{(G)}q_f^{(H)} \neq 0,
\end{equation}
where the summation is taken over all massless fields $\Psi '_0$
which are available below the scale $M_I$.
Carrying out the orthogonal transformation for gauge fields again
\begin{eqnarray}
                               \left(
   \begin{array}{l}
        K_{\mu } \\
        L_{\mu }
   \end{array}
                               \right)
                            & = &
                               \left(
   \begin{array}{cc}
        \cos \phi & \sin \phi \\
       -\sin \phi & \cos \phi
   \end{array}
                               \right)
                               \left(
   \begin{array}{l}
        G_{\mu } \\
        H_{\mu }
   \end{array}
                               \right), \nonumber \\
                               \left(
   \begin{array}{l}
      g_K(M_I)Q_K \\
      g_L(M_I)Q_L
   \end{array}
                               \right)
                            & = &
                               \left(
   \begin{array}{cc}
      \cos \phi & \sin \phi \\
     -\sin \phi & \cos \phi
   \end{array}
                               \right)
                               \left(
   \begin{array}{l}
      g_G(M_I)Q_G \\
      g_H(M_I)Q_H
   \end{array}
                               \right)
\end{eqnarray}
with
\begin{equation}
    \tan 2\phi = \frac {2g_G(M_I)\, g_H(M_I)\, b'_{GH}}
                        {g_G^2(M_I)\, b'_{GG}-g_H^2(M_I)\, b'_{HH}}\ ,
\end{equation}
we obtain
\begin{eqnarray}
    b'_{KL}           &=& 0,        \nonumber \\
   g_K^2(M_I) b'_{KK} &=&
         \frac{1}{2}\left\{ g_G^2(M_I)\, b'_{GG}
                            +g_H^2(M_I)\, b'_{HH} \right\}  \nonumber \\
      & & \makebox[5em]{}   +\frac{g_G^2(M_I)\, b'_{GG}
                             -g_H^2(M_I)\, b'_{HH}}{2\cos 2\phi }, \\
   g_L^2(M_I) b'_{LL} &=&
         \frac{1}{2}\left\{ g_G^2(M_I)\, b'_{GG}
                             +g_H^2(M_I)\, b'_{HH} \right\}  \nonumber \\
      & & \makebox[5em]{}    -\frac{g_G^2(M_I)\, b'_{GG}
                              -g_H^2(M_I)\, b'_{HH}}{2\cos 2\phi }.
                                                               \nonumber
\end{eqnarray}
The gauge couplings $g_G(M_I)$ and $g_H(M_I)$ at the scale $M_I$
do not necessarily take the same values.
The new gauge couplings $g_K(M_I)$ and $g_L(M_I)$ are given as
\begin{eqnarray}
   g_K^2(M_I) &=& g_G^2(M_I) \cos ^2\phi +g_H^2(M_I) \sin ^2\phi ,
                                                          \nonumber \\
   g_L^2(M_I) &=& g_G^2(M_I) \sin ^2\phi +g_H^2(M_I) \cos ^2\phi ,
\label{eqn:gc1}
\end{eqnarray}
respectively.
The one-loop renormalization group equations read
\begin{equation}
   \frac{d\,g_i}{d\,\ln \mu}=\frac{b'_{ii}}{4\pi }g_i^3
          \ \ \ \ \ \ \ (i=K,L)
\end{equation}
in the region below $M_I$.

\section{Four-Generation Models}
\hspace*{\parindent}
Let us consider four-generation
superstring models which are obtained through
the Calabi-\-Yau compactification with $h^{11}=1$ and $h^{21}=5$
\cite{4generation}.
The manifold $K$ considered here is non-simply connected and
constructed by moding $K_0$ by a discrete symmetry group $G_d$ of $K_0$
as $K=K_0/G_d$.
$K_0$ is a simply connected Calabi-Yau manifold defined as
a hypersurface in $CP^4$ with the specific defining polynomial
$\sum_iz_i^5=0$, where $z_i(i=1\sim 5)$ are homogeneous coordinates
in $CP^4$.
This manifold $K_0$ has the high discrete symmetry $S_5\times Z_5^5/Z_5$.
The discrete group $G_d$ is taken as $Z_5\times Z_5$
which is a subgroup of the high discrete symmetry.

As a consequence of the high discrete
symmetry of $K_0$ we are led to introduce
a large intermediate energy scale $M_I$, which is defined by VEVs of
$SU(3)_c\times SU(2)_L\times U(1)_Y$-neutral
fields $S$ and $\overline {S}$
belonging to ${\bf 27}$ and $\overline {\bf 27}$ in $E_6$, respectively.
More explicitly, the large intermediate scale $M_I$ is given by
\cite{4generation}
\begin{equation}
   M_I=\langle S\rangle =\langle \overline {S}\rangle
       \sim M_{NR}\left( \frac {M_S}{M_{NR}}\right)^{1/6}.
\end{equation}
Here $M_S$ stands for the soft supersymmetry breaking scale and
is taken as $M_S\sim 10^3$GeV.
Since the $D$-flatness is guaranteed by the equality $\langle S\rangle
=\langle \overline S\rangle $,
the supersymmetry is unbroken at the intermediate scale $M_I$.
The mass scale $M_{NR}$, which appears as its inverse power
in the non-renormalizable terms of the superpotential, becomes~
\cite{Kalara}
\begin{equation}
    M_{NR} \sim M_C\,|\eta (iT)|^{-2} >M_C\,.
\end{equation}
Thus we get the large intermediate scale $M_I \sim 10^{16}$GeV.
Since leptoquark particles can gain masses of order of $M_I$,
this large value of $M_I$ is consistent with the proton stability.
The four-generation superstring models are candidates of viable models
which reproduce the standard model at low energies.

Due to the flux breaking mechanism the discrete group
$G_d$ is embedded into $E_6$ and
the gauge group $G$ at the string scale
prevailingly becomes smaller than $E_6$.
We denote the embedding of $G_d$ into $E_6$ as $\overline {G_d}$.
If and only if $\overline {G_d}$ is taken as $Z_5$, we obtain the following
two types of realistic gauge hierarchies
\cite{Matsuoka}
\begin{description}
   \item [\ \ \ \ (i)]
   $G=SU(3)_C\times SU(2)_L\times SU(2)_R\times U(1)\times U(1)$,
   \item [\ \ \ \ (ii)] $G=SU(4)_C\times SU(2)_L\times U(1)\times U(1)$.
\end{description}
Matter contents in the models are different from one to another.
This difference is of critical importance in the evolution of
gauge coupling constants.
In Table 1 are shown chiral superfields, their representations
in $G$ and their multiplicities.
As seen in Table 1, it is noted that $SU(4)_C$ in the model (ii)
is in contrast to that in Pati-Salam model.
\bigskip
\begin{center}
          {\large {\bf Table 1}}
\end{center}
\bigskip

At the intermediate scale $M_I$ the gauge symmetry $G$ is
spontaneously broken into the smaller gauge group $G'$
via the non-vanishing VEVs of $S$ and $\overline S$.
In each model the remaining gauge group $G' $ becomes
\begin{description}
   \item[\ \ \ \ (i)]  $G' =SU(3)_C\times SU(2)_L\times
       SU(2)_R\times U(1)_{B-L}$,
   \item[\ \ \ \ (ii)] $G' =SU(3)_C\times SU(2)_L\times
       U(1)_Y\times U(1)_{\chi }$,
\end{description}
where $U(1)_{\chi }$ stands for $SO(10)/SU(5)$.
When the fields $S$ and $\overline S$ develop non-zero VEVs,
the leptoquark fields $g$, $g^c$ and Higgs fields $H_u, H_d
(\overline {H_u}, \overline {H_d})$ can get masses of order $M_I$
through the Yukawa interactions $gg^cS$ and $H_uH_dS$.
In the present analysis it is assumed that all of $g$, $g^c$,
$H_u$ and $H_d$ but only one family of
$H_u$ and $H_d$ gain masses of $O(M_I)$.
On the other hand, all of the matter fields which can not couple
with $S$ through the Yukawa
interactions remain massless at the scale $M_I$.

Here the gauge symmetry $G'$ is assumed to be spontaneously broken
to the standard gauge group via a non-zero VEV of sneutrino
at the energy scale $M_R$.
Since the magnitude of VEVs of $F$-terms and $D$-terms are limited by
the soft susy breaking scale $M_S^2$,
$M_R$ should be the same order with $M_S$.
Thus hereafter we take $M_R=M_S$.

Through analysis of non-abelian gauge couplings in Ref.
\cite {Tsunoda},
it was found that the model (i) is unfavorable for the string unification,
while the model (ii) is consistent with the unification condition (9).
Main difference between two models comes from extra $d^c$ and
$\overline {d^c}$ contribution in the region  ranging from $M_I$ to $M_S$.
Therefore, in the next section we focus our attention
to the string unification in the model (ii).
The running of gauge couplings including abelian gauge couplings
is systematically studied.
In Table 2 we tabulate multiplicities of matter fields in
the respective energy ranges derived from the above scenarios.
\bigskip
\begin{center}
           {\large {\bf Table 2}}
\end{center}
\bigskip

\section
{Renormalization Group Analysis}
\hspace*{\parindent}

Now we investigate the evolution of gauge couplings in the model (ii)
and whether the gauge unification occurs or not.
The gauge group $G$ at the scale $M_C$ is
\begin{equation}
   G=SU(4)_C\times SU(2)_L\times U(1)_{\gamma }\times U(1)_{\delta }.
\end{equation}
$U(1)_{\gamma }$-charge is given by a linear combination of
$U(1)_Y$- and $U(1)_{\eta }$-charges as
\begin{eqnarray}
    Q_{\gamma } &=& \frac {3}{\sqrt {10}}\,Q_Y
                        +\frac {1}{\sqrt {10}}\,Q_{\eta }, \\
    Q_{\eta }   &=& \sqrt {\frac {5}{8}}\,Q_{\psi }
                        -\sqrt {\frac {3}{8}}\,Q_{\chi },
\end{eqnarray}
where $U(1)_{\psi }$ stands for $E_6/SO(10)$.
$U(1)_{\delta }$-axis coincides with one of the root vector of $E_6$
and is expressed as
\begin{equation}
    Q_{\delta }= -\sqrt {\frac {3}{8}}\,Q_{\psi }
                     -\sqrt {\frac {5}{8}}\,Q_{\chi }.
\end{equation}
$U(1)_{\delta }$-axis is perpendicular to
$U(1)_Y$- and $U(1)_{\eta }$-axes.
The charges for matter fields are shown in Table 3.
{}From this Table we find
\begin{eqnarray}
   b_{\gamma \gamma } &=& 3N_f+\frac{1}{3}=\frac{37}{3}, \nonumber \\
   b_{\delta \delta } &=& 3N_f+2=14, \\
   b_{\gamma \delta } &=& -\frac{2}{\sqrt 6}. \nonumber
\end{eqnarray}
\bigskip
\begin{center}
     {\large {\bf Table 3}}
\end{center}
\bigskip
Since $b_{\gamma \delta }\neq 0$, we consider the transformation
of the $U(1)$-basis in such a way that off-diagonal elements
of $b_{ij}$ become vanishing.
{}From Eq.(\ref{eqn:omega}) the rotation angle $\omega $
is determined by
\begin{equation}
   \tan 2\omega = \frac{2\sqrt 6}{5}.
\end{equation}
Thus we fix the transformation from
$U(1)_{\gamma }\times U(1)_{\delta }$
to $U(1)_{\theta }\times U(1)_{\xi }$
\begin{eqnarray}
   Q_{\theta} & = & \sqrt{6\over7}\,
   Q_{\gamma }+\sqrt{1\over7}\,Q_{\delta },
                            \nonumber \\
   Q_{\xi }   & = &-\sqrt{1\over7}\,
   Q_{\gamma }+\sqrt{6\over7}\,Q_{\delta }.
   \label{eqn:PQAB}
\end{eqnarray}
New $U(1)$-charges are also tabulated in Table 3.

In the energy range $M_C\geq \mu \geq M_I$,
the one-loop evolution of gauge couplings is expressed as
\begin{equation}
   \alpha^{-1}_i(\mu) = \alpha^{-1}_{st} +\frac {b_{ii}^{[1]}}{2\pi}
                        \ln\left(
 	                         	{{M_X}^2\over \mu^2}
                           \right),
   \label{eqn:TTHRES}
\end{equation}
where $i=4,2,\theta ,\xi $ stand for $SU(4)_C$, $SU(2)_L$,
$U(1)_{\theta }$ and $U(1)_{\xi }$ components, respectively.
The one-loop coefficients of $\beta$-function in this region become
\begin{equation}
   b_{ij}^{[1]}={\rm diag}
   	\pmatrix{ 1  &  6  &  12  &  43/3 }\ \ \ \ \
        (i,j=4,2,\theta ,\xi ) .
   \label{eqn:BXI}
\end{equation}
Up to two-loop level the renormalization group equations
for gauge couplings has the form
\cite {Jones}
\begin{equation}
   \frac {d g_i}{d\ln\mu}  =
                \sum_j \left\{ \frac {b_{ij}^{[1]}}{4\pi}g_j^2\,g_i
                +{b_{ij}^{[2]}\over (4\pi)^2}g_j^2\,g_i^3 \right\}
   +\sum_{j,k\neq i}{{b}_{ijk}^{[2]}\over 2(4\pi)^2}g_k^2\,g_j^2\,g_i.
\label{eqn:twoloop}
\end{equation}
In order to distinguish between the one-loop contribution
and the two-loop one, here we attach superscript $[1]$ and $[2]$
to $b$'s.
The two-loop coefficients of $\beta$-function are given by
\begin{equation}
   b_{ij}^{[2]}=\pmatrix{  231/2  &   12   &   4   &   31/6   \cr
                             60   &   60   &   4   &    4     \cr
                             60   &   12   &  12   &    4     \cr
                           155/2  &   12   &   4   &  265/18  }
                   \ \ (i,j=4,2,\theta ,\xi ).
\end{equation}
In the following analysis the last term in Eq.(\ref{eqn:twoloop})
can be neglected numerically because of its smallness.

In the present model the symmetry breaking occurs
at the scale $M_I \sim 10^{16}$GeV
due to nonzero VEVs of $S$ and $\overline S$.
The gauge symmetry is spontaneously broken as
\begin{equation}
   SU(4)_C\times U(1)_{\xi } \rightarrow SU(3)_C\times U(1)_{\rho },
\end{equation}
whereas $SU(2)_L\times U(1)_{\theta }$ remains unbroken.
This $U(1)_{\rho }$ gauge field $V_{\mu}^{(\rho )}$,
its generator $Q_{\rho }$  and its gauge coupling
$g_{\rho }(M_I)$ are given by
\begin{eqnarray}
   V_{\mu }^{\rho } & = & \frac{1}{\sqrt{9g_4(M_I)^2+7g_{\xi }(M_I)^2}}
                           \left\{ \sqrt{7}g_{\xi }(M_I)\,G_{\mu}
                                   +3g_4(M_I)\,V_{\mu}^{(\xi)} \right\}, \\
   Q_{\rho }      & = & \frac {1}{4}(\sqrt 7 \;T_{15}+3Q_{\xi }), \\
   g_{\rho }(M_I) & = & \frac {4g_4(M_I)\, g_{\xi }(M_I)}
                         {\sqrt{9g_4(M_I)^2+7g_{\xi }(M_I)^2}},
\end{eqnarray}
where $G_{\mu }$ stands for the gauge field associated with $T_{15}$ and
$g_4(M_I)$ and $g_{\xi }(M_I)$ are the gauge couplings at the scale $M_I$
for $SU(4)_C$ and $U(1)_{\xi }$, respectively.
A generator $T_{15}$ of $SU(4)_C$ is of the form
$T_{15}={\rm diag}(1,1,1,-3)/2\sqrt 6$
for the fundamental representation.
At the intermediate scale $M_I$ we have two $U(1)$ gauge groups
i.e. $U(1)_{\theta }\times U(1)_{\rho }$.
In Table 4 we show these $U(1)$-charges of matter fields.
Through the symmetry breaking all of $g$, $g^c$, $H_u$ and $H_d$
but only one family of $H_u$ and $H_d$ become massive at $M_I$.
These massive fields do not contribute to the evolution of
gauge couplings at energies below $M_I$.
Since the summation runs over all massless fields $\Psi '_0$, we obtain
\begin{eqnarray}
  {b'_{\theta \rho }}^{[1]} &=&
                \sum _{f\in \Psi '_0} q_f^{(\theta )}q_f^{(\rho )}
                     =-\frac {\sqrt 6}{7} \neq 0, \nonumber \\
  {b'_{\theta \theta }}^{[1]} &=&
                \sum _{f\in \Psi '_0} \left(q_f^{(\theta )}\right)^2
                     =\frac {57}{7}, \\
  {b'_{\rho \rho }}^{[1]} &=&
                \sum _{f\in \Psi '_0} \left(q_f^{(\rho )}\right)^2
                     =\frac {297}{28}.  \nonumber
\end{eqnarray}
\bigskip
\begin{center}
   {\large {\bf Table 4}}
\end{center}
\bigskip
As explained in section 2, we carry out the orthogonal transformation
for $U(1)$ gauge fields.
If we denote new $U(1)$-basis as $U(1)_{\sigma }\times U(1)_{\tau }$,
we get
\begin{eqnarray}
                                      \left(
   \begin{array}{l}
        V^{(\sigma )}_{\mu } \\
        V^{(\tau )}_{\mu }
   \end{array}
                                      \right)
                                  & = &
                                      \left(
   \begin{array}{cc}
        \cos \phi & \sin \phi \\
       -\sin \phi & \cos \phi
   \end{array}
                                      \right)
                                      \left(
   \begin{array}{l}
        V^{(\theta )}_{\mu } \\
        V^{(\rho )}_{\mu }
   \end{array}
                                      \right), \nonumber  \\
                                      \left(
   \begin{array}{l}
        g_{\sigma }(M_I)\,Q_{\sigma } \\
        g_{\tau }(M_I)\,  Q_{\tau }
   \end{array}
                                      \right)
                                  & = &
                                      \left(
   \begin{array}{cc}
        \cos \phi & \sin \phi \\
       -\sin \phi & \cos \phi
   \end{array}
                                      \right)
                                      \left(
   \begin{array}{l}
        g_{\theta }(M_I)\,Q_{\theta } \\
        g_{\rho }(M_I)\,  Q_{\rho }
   \end{array}
                                      \right)
\end{eqnarray}
with the rotation angle
\begin{equation}
   \tan 2\phi =
             \frac{2g_{\theta}(M_I)\, g_{\rho}(M_I)\,
             		{b'_{\theta \rho}}^{[1]}}
                  {g_{\theta }(M_I)^2\, {b'_{\theta \theta}}^{[1]}
                        -g_{\rho }(M_I)^2\,{b'_{\rho \rho}}^{[1]}} \ .
\end{equation}
The gauge couplings of new $U(1)$-basis are given by
\begin{eqnarray}
  g_{\sigma }(M_I)^2 &=& g_{\theta }(M_I)^2 \cos ^2\phi
                         +g_{\rho }(M_I)^2 \sin ^2\phi ,
                              \nonumber \\
  g_{\tau }(M_I)^2   &=& g_{\theta }(M_I)^2 \sin ^2\phi
                         +g_{\rho }(M_I)^2 \cos ^2\phi .
\end{eqnarray}

In the range $M_I - M_S$
the one-loop coefficients of the $\beta $-function become
\begin{equation}
       {b'_{ij}}^{[1]} = {\rm diag}
              \pmatrix{ 0  &  3  &  {b'_{\sigma \sigma}}^{[1]}
                                  &  {b'_{\tau \tau }}^{[1]}) }
              \ \ \ \ \ (i,j=3,2,\sigma ,\tau ),
\end{equation}
where
\begin{eqnarray}
   g_{\sigma }(M_I)^2\,{b'_{\sigma \sigma }}^{[1]} &=&
      \frac {1}{2}\left\{ g_{\theta }(M_I)^2\, {b'_{\theta \theta }}^{[1]}
            +g_{\rho }(M_I)^2\,
            {b'_{\rho \rho }}^{[1]}\right\}  \nonumber \\
       & & \makebox[5em]{}   +\frac{ g_{\theta }(M_I)^2\,
                                  {b'_{\theta \theta }}^{[1]}
            -g_{\rho }(M_I)^2\, {b'_{\rho \rho }}^{[1]}}
         {2 \cos 2\phi },                             \nonumber \\
   g_{\tau }(M_I)^2\,{b'_{\tau \tau }}^{[1]} &=&
       \frac {1}{2}\left\{ g_{\theta }(M_I)^2\, {b'_{\theta \theta }}^{[1]}
             +g_{\rho }(M_I)^2\,
             {b'_{\rho \rho }}^{[1]}\right\}  \nonumber \\
       & & \makebox[5em]{}    -\frac{g_{\theta }(M_I)^2\,
                                   {b'_{\theta \theta }}^{[1]}
         -g_{\rho }(M_I)^2\, {b'_{\rho \rho }}^{[1]}}
         {2 \cos 2\phi }.
\end{eqnarray}
If the gauge kinetic mixing between $U(1)_{\theta }$ and $U(1)_{\rho }$
is negligible at one-loop level,
the two-loop coefficients of $\beta$ function in this region
could be expressed as
\begin{equation}
   {b'_{ij}}^{[2]} =  \pmatrix{
                                48   &   12   &   12/7   &   67/14    \cr
                                32   &   39   &   25/7   &   10/7     \cr
                               96/7  &  75/7  &  489/49  &  162/49    \cr
                              268/7  &  30/7  &  162/49  & 4619/392
                             } \ \ (i,j=3,2,\theta ,\rho ).
\end{equation}
Since $U(1)_{\sigma }$- and $U(1)_{\tau }$-charges are functions of
$g_{\theta }(M_I)$, $g_{\rho }(M_I)$ and $\phi $,
the expressions in terms of $U(1)_{\sigma }\times U(1)_{\tau }$-basis
are complicated and
the correct ${b'_{ij}}^{[2]}$ is obtained numerically as seen later.
Thus we can solve the renormalization group
equations over the energy range
$M_I \geq \mu \geq M_S$.

The next stage of the symmetry breaking is brought about
at the scale $M_R=M_S \sim 10^3$GeV through
a nonzero VEV of sneutrino $\tilde{\nu ^c}$.
The gauge symmetry $G'=SU(3)_C\times SU(2)_L\times U(1)_{\sigma }
\times U(1)_{\tau }$ is spontaneously broken into
$G_{st}=SU(3)_C\times SU(2)_L\times U(1)_Y$.
Then we have the relation
\begin{eqnarray}
   Q_Y & = & \sqrt {\frac{27}{35}}\,Q_{\theta }
                            -\sqrt {\frac{8}{35}}\,Q_{\rho }, \\
   g_Y(M_S)\,Q_Y & = & g_{\sigma }(M_S)\, Q_{\sigma }\cos \varphi
                       +g_{\tau }(M_S)\, Q_{\tau }\sin \varphi.
\end{eqnarray}
The angle $\varphi $ is defined by
\begin{equation}
   \tan \varphi = -\frac {g_{\sigma }(M_S)\,u}{g_{\tau }(M_S)\,v},
\end{equation}
where
\begin{eqnarray}
     u &=& \frac {1} {g_{\sigma }(M_I)}  \left(
               \sqrt{\frac{8}{35}}\, g_{\theta }(M_I)\cos \phi
              +\sqrt{\frac{27}{35}}\, g_{\rho }(M_I)\sin \phi \right),
                                            \nonumber  \\
     v &=& \frac {1} {g_{\tau }(M_I)}  \left(
              -\sqrt{\frac{8}{35}}\,g_{\theta }(M_I)\sin \phi
              +\sqrt{\frac{27}{35}}\,g_{\rho }(M_I)\cos \phi \right).
\end{eqnarray}
The $U(1)_Y$-gauge coupling is of the form
\begin{equation}
     g_Y(M_S) = \frac{g_{\sigma}(M_S)\,g_{\tau}(M_S)}
                     {\sqrt{\{g_{\sigma}(M_S)\,u\}^2
                           +\{g_{\tau}(M_S)\,v\}^2    }}
                \frac{g_{\theta}(M_I)\,g_{\rho}(M_I)}
                     {g_{\sigma}(M_I)\,g_{\tau}(M_I)}.
\end{equation}
According to the symmetry breaking,
$\nu ^c$ and the 4-th generation of $l$ and $e^c$
decouple from the effective theory at energies below $M_S$.
Detailed mass spectra of neutralinos and charginos around
this scale have been studied in Ref.%
\cite {CHARGINO}.
Finally, at energies below $M_S$ the effective theory
becomes non-supersymmetric.
As a consequence, in the region from $M_S$ to $M_Z$
the evolution of gauge couplings is descibed in terms of the
coefficients of the $\beta $-function
\begin{eqnarray}
   {b''_{ij}}^{[1]} & = & {\rm diag}
   \pmatrix{-17/3  &  -2  &  74/15 }, \\
   {b''_{ij}}^{[2]} & = & \pmatrix{  -2/3  &    6   &   22/15   \cr
                                  16   &  81/4  &    5/4    \cr
                                176/15 &  15/4  &  277/60   \cr
                          }\quad (i,j=3,2,Y).
\end{eqnarray}
In Table 5 we illustrate gauge hierarchies of the present model.
\bigskip
\begin{center}
   {\large {\bf Table 5}}
\end{center}
\bigskip

We are now in a position to carry out numerical analysis of
the renormalization group evolution of gauge couplings.
In the four-generation Calabi-Yau models the intermediate energy
scale $M_I$ is around $10^{16}$GeV.
Therefore, here we take $M_I$ as
\begin{equation}
    M_I=10^{15\sim 17}{\rm GeV}.
\end{equation}
And also the soft supersymmetry breaking scale $M_S$ is taken as
\begin{equation}
    M_S=10^{2\sim 4}{\rm GeV}.
\end{equation}
In the present analysis, after studying the solution
in which unification condition
is fulfilled for non-abelian gauge couplings, we manipulate
the evolution of the abelian gauge couplings.
As a consequence of the contribution
from extra $d^c$ and ${\overline d^c}$
fields in the energy range $M_I-M_S$, the unification of nonabelian
gauge couplings occurs at $\alpha_{st}^{-1}=6\sim 10$
and $M_X=O(10^{19})$GeV.
Taking this situation into account, let us consider the following
two parametrizations for a universal
value of gauge coupling and energy scales. \\
Case (a):
\begin{eqnarray}
     \alpha_{st}^{-1} = 6.09, \ \ \ \ \ \ &
                     M_X = 1.1\times 10^{19}\gev, \nonumber \\
           M_I = 1.0\times 10^{16}\gev,  & M_S = 1.0\times 10^3\gev.
\end{eqnarray}
Case (b):
\begin{eqnarray}
     \alpha_{st}^{-1} = 6.14, \ \ \ \ \ \ &
                     M_X = 3.3\times 10^{18}\gev, \nonumber \\
           M_I = 1.0\times 10^{15}\gev,  & M_S = 1.0\times 10^3\gev.
\end{eqnarray}

As numerical results in the case (a) we have
\begin{equation}
      M_C=1.01\times 10^{18}\gev
\end{equation}
and the magnitudes of the string threshold corrections
\begin{eqnarray}
     \alpha_4^{-1}(M_C)
     -\alpha_{st}^{-1} & = & 1.00, \nonumber  \\
     \alpha_2^{-1}(M_C)
     -\alpha_{st}^{-1} & = & 2.84, \nonumber  \\
     \alpha_{\theta }^{-1}(M_C)
     -\alpha_{st}^{-1} & = & 4.96,            \\
     \alpha_{\xi }^{-1}(M_C)
     -\alpha_{st}^{-1} & = & 5.94. \nonumber
\end{eqnarray}
It is worthy to note that the threshold corrections
are sizable compared with $\alpha _{st}^{-1}$.
By the use of Eqs.(34) to (36),
the running gauge couplings in the region $M_C-M_I$ can be calculated.
The gauge couplings at the intermediate scale $M_I$ become
\begin{eqnarray}
     \alpha_4^{-1}(M_I)          & = &  8.76,  \nonumber  \\
     \alpha_2^{-1}(M_I)          & = & 14.11,  \nonumber  \\
     \alpha_{\theta }^{-1}(M_I)  & = & 20.41,             \\
     \alpha_{\xi }^{-1}(M_I)     & = & 23.24.  \nonumber
\end{eqnarray}
Using Eq.(40) we have
\begin{equation}
     \alpha_{\rho }^{-1}(M_I) = 16.90.
\end{equation}
{}From Eq.(44), in the $SU(3)_C\times SU(2)_L\times
U(1)_{\sigma }\times U(1)_{\tau }$-basis these results
are expressed as
\begin{eqnarray}
     \alpha_3^{-1}(M_I)           & = &  8.76,  \nonumber  \\
     \alpha_2^{-1}(M_I)           & = & 14.11,  \nonumber  \\
     \alpha_{\sigma }^{-1}(M_I)   & = & 20.38,             \\
     \alpha_{\tau }^{-1}(M_I)     & = & 16.92,  \nonumber
\end{eqnarray}
where the rotation angle $\phi $ is given by
\begin{equation}
            \tan \phi = 0.082.
\end{equation}
Next, the coefficients of
the $\beta $-function in the region $M_I-M_S$
are numerically estimated as
\begin{eqnarray}
       {b'_{ij}}^{[1]} & = & {\rm diag}
                  \pmatrix{
                        0.00  &  3.00  &  8.10  &  10.65
                              },       \nonumber \\
   {b'_{ij}}^{[2]} & = & \pmatrix{
                     48.00   &   12.00   &   1.57   &   4.91    \cr
                     32.00   &   39.00   &   3.74   &   1.28    \cr
                     12.58   &   11.22   &   9.75   &   3.55    \cr
                     39.26   &    3.85   &   3.55   &  11.54
                             }. \ \ (i,j=3,2,\sigma ,\tau )
\end{eqnarray}
Due to the gauge kinetic mixing
the elements of ${b'_{ij}}^{[2]}$ with $i$ and/or $j=\sigma ,\tau $
deviate from those in Eq.(47) about 10\%.
After running the gauge couplings down
to the soft susy breaking scale $M_S$,
we obtain
\begin{eqnarray}
     \alpha_3^{-1}(M_S)          & = & 10.89, \nonumber  \\
     \alpha_2^{-1}(M_S)          & = & 30.38, \nonumber  \\
     \alpha_{\sigma }^{-1}(M_S)  & = & 59.79,            \\
     \alpha_{\tau }^{-1}(M_S)    & = & 69.37.  \nonumber
\end{eqnarray}
Using Eq.(52) we get
\begin{equation}
     \alpha_Y^{-1}(M_S) =  63.60
\end{equation}
with $\tan \varphi =-0.711$.
Finally the gauge couplings at the electroweak scale turn out to be
\begin{eqnarray}
     \alpha_3^{-1}(M_Z) & = & 8.82,  \nonumber  \\
     \alpha_2^{-1}(M_Z) & = & 29.71,            \\
     \alpha_Y^{-1}(M_Z) & = & 65.45. \nonumber
\end{eqnarray}
\bigskip
\begin{center}
{\large {\bf Fig.2}}
\end{center}
\bigskip
In Fig.2 we show the running behavior
of the gauge couplings for the case (a).
On the other hand, experimental values of gauge couplings
at the electroweak scale $M_Z$ are
\cite {PDG}
\begin{eqnarray}
    \alpha_3^{-1}(M_Z) & = &  8.83 \pm 0.28, \nonumber  \\
    \alpha_2^{-1}(M_Z) & = & 29.75 \pm 0.11,            \\
    \alpha_Y^{-1}(M_Z) & = & 58.89 \pm 0.11. \nonumber
\end{eqnarray}
The calculated value of $\alpha _Y^{-1}(M_Z)$ is about 10\% too large
compared with the experimental value.

To improve the calculated value of $\alpha _Y^{-1}(M_Z)$,
it seems to be better to lower the intermediate scale
while to fix the soft susy breaking scale.
Thus we take the case (b) as the second parametrization.
In the case (b) we obtain
\begin{equation}
       M_C=1.00\times 10^{18}\gev
\end{equation}
and
\begin{eqnarray}
     \alpha_3^{-1}(M_Z) & = &  9.01, \nonumber  \\
     \alpha_2^{-1}(M_Z) & = & 29.65,            \\
     \alpha_Y^{-1}(M_Z) & = & 64.16. \nonumber
\end{eqnarray}
The value of $\alpha _Y^{-1}$ is still inconsistent with the
experimental value.
The other para\-metriza\-tions also can hardly improve
the situation significantly.
In conclusion, it is difficult to unify
the gauge couplings along this scenario.

\section{Summary and Discussion}
\hspace*{\parindent}

We investigated the unification of gauge couplings in the Calabi-Yau
superstring model.
In these models with one K\"ahler class modulus the unification
scheme is constrained by the string threshold corrections.
This constraint implies that gauge couplings join at the energy
larger than the string compactification scale, i.e. $M_X > M_C$.
On the other hand, in the renormalization group evolution of
abelian gauge couplings we are frequently confronted with
a gauge kinetic mixing problem.
In superstring models the rank of the gauge group
at $M_C$ is larger than that of the standard gauge group.
Then in many of the superstring derived models more than one $U(1)$
gauge group are contained at various stages of symmetry breaking
and there possibly exist gauge kinetic mixing in the effective theory.
In fact, as a consequence of the contribution from extra fields
which do not exist in the MSSM,
it was found that the gauge kinetic mixing
takes place in the four-generation
superstring models.
In the presence of such mixing it is necessary for us to
diagonalize the gauge kinetic terms.
We studied the mixing effect in
the four-generation Calabi-Yau model systematically through this paper.
At the string scale the $U(1)$-basis was chosen in such a way
that the gauge kinetic mixing terms disappear.
At energies below the intermediate energy scale $M_I$
at which the gauge symmetry is spontaneously broken,
the gauge kinetic mixing terms emerge again in the effective theory.
Then we carried out the transformation of $U(1)$ -basis again
and selected the basis in which we have no gauge kinetic mixing.
By solving the renormalization group equations for
gauge couplings, we explored an advisable solution which is
consistent with the experimental values of gauge
couplings at the electroweak scale $M_Z$.
However, no such a solution is found.
When we first take a desirable solution for the unification
of non-abelian gauge couplings,
the extrapolated value of $\alpha _Y^{-1}$ from the string scale
to the electroweak scale is $\sim 10$\% too large compared with
the experimental value.

{}From the viewpoint of the string unification the constraints
on the energy scale and matter contents
are quite instructive to construct viable superstring models.
One of the interesting possibilities is the case that
a right-handed sneutrino has a large VEV such as $10^{10\sim 12}$GeV.
To preserve the supersymmetry at this scale,
the conditions of the F-flatness and the D-flatness should
be guaranteed.
Then in such models we need to have at least a pair of $\nu ^c$ of
${\bf 27}$ and $\overline {\nu ^c}$ of ${\overline {\bf 27}}$.
As an example of the models in which $\nu ^c$ and $\overline {\nu ^c}$
as well as $S$ and $\overline S$ are contained as chiral superfields,
the three-generation model is known
\cite {3gener}.
In the three-generation model, however,
there appear too many mirror quarks and leptons.
Then, when we carry out the evolution of gauge couplings
from the scale $M_Z$ to higher energy scale,
the gauge couplings blow up at energies below $10^{15}$GeV
\cite {Dela}.
In addition, the intermediate scale $M_I$ turns out to be
lower than $O(10^{15})$GeV in the model.
This contradicts with the proton stability.
Making use of the constraints from the string unification
on the energy scale and matter contents,
it is interesting to find out viable models
in which we have $S$, $\nu ^c$ and $\overline S$, $\overline {\nu ^c}$
chiral superfields.

\newpage

{\large {\bf Table Captions}}
\bigskip

{\bf Table 1} \ \ \ \ \ \ Chiral superfields belonging to ${\bf 27}$ and
${\overline {\bf 27}}$ in $E_6$.
These fields are classified according as their $Z_5$-charges of
the discrete group ${\overline G_d}$ which is an embedding
of $G_d$ into $E_6$.
Their representations in the non-abelian gauge groups
$SU(3)_C\times SU(2)_L\times SU(2)_R$ for the model (i) and
$SU(4)_C\times SU(2)_L$ for the model (ii) are described
in the parentheses.
Their multiplicities
( their generation numbers and anti-generation numbers)
at the string scale are also given in the square brackets.
\bigskip

{\bf Table 2} \ \ \ \ \ \
Multiplicities(generation numbers) of available fields
in the respective energy ranges $M_C - M_I$, $M_I - M_S$ and
$M_S - M_Z$.
\bigskip

{\bf Table 3} \ \ \ \ \ \ Quantum numbers
of chiral superfields in the bases
used at $M_C$ and in the range $M_C - M_I$.
The representations of chiral superfields in $SU(4)_C\times SU(2)_L$
are also shown in the parentheses of the first row.
To diagonalize $b_{ij}$ $U(1)_{\gamma }\times U(1)_{\delta }$-basis
is transformed to $U(1)_{\theta }\times U(1)_{\xi }$-basis (see text).
\bigskip

{\bf Table 4} \ \ \ \ \ \ $U(1)$-charges of chiral superfields
in the bases used at $M_I$ and $M_S$.
The spontaneous breaking $SU(4)_C\times U(1)_{\xi } \rightarrow
SU(3)_C\times U(1)_{\rho }$ occurs at the scale $M_I$.
\bigskip

{\bf Table 5} \ \ \ \ \ \ Gauge hierarchies
in the present four-generation model.
At the string scale $M_C$ and at the intermediate scale $M_I$
the transformations of $U(1)$-basis are carried out so as to
diagonalize the one-loop coefficients of the $\beta $-function.

\newpage

{\large {\bf Figure Captions}}
\bigskip

{\bf Fig. 1} \ \ \ \ \ \ The gauge kinetic mixing at one-loop level.
When $b_{AB} \neq 0$, this diagram gives rise to the mixing term
$F_{\mu \nu }^{(A)}{F^{(B)}}^{\mu \nu }$.
\bigskip

{\bf Fig. 2} \ \ \ \ \ \ Evolution of the gauge couplings in the case (a).
The intermediate scale $M_I$ and the soft susy breaking scale $M_S$
are taken as $10^{16}$GeV and $10^3$GeV, respectively.
The energy scale $M_X$ at which gauge couplings have a common value
is $1.1\times 10^{19}$GeV.
The experimental values of gauge couplings at the electroweak scale
are also indicated.

\newpage

\renewcommand{\arraystretch}{1.5}

\begin{center}
{\bf Table 1} \\
\bigskip

\begin{tabular}{|c|c|lr|lr|} \hline
model & $Z_5$-charge &
  \multicolumn{2}{|c|}{fields in {\bf 27} } &
   \multicolumn{2}{|c|}{fields in ${\overline {\bf 27}}$ } \\
     \hline \hline
 & 0 & $S$       & (1,1,1)
 & ${\overline S}$                   & (1,1,1) \\
 &   & $H_u,H_d$ & (1,2,2)
 & ${\overline H_u},{\overline H_d}$ & (1,2,2) \\
 &   & \multicolumn{2}{|c|}{[5 generations]} &
       \multicolumn{2}{|c|}{[1 anti-generation]} \\
\cline{2-6}
 & 1 & $d^c,u^c$ & ($3^*$,1,2) & \multicolumn{2}{|c|}{------} \\
 &   & \multicolumn{2}{|c|}{[4 generations]} & \multicolumn{2}{|c|}{} \\
\cline{2-6}
    & 2 & $g$       & (3,1,1)     & \multicolumn{2}{|c|}{------} \\
(i) &   & $e^c,\nu ^c$& (1,1,2)   & \multicolumn{2}{|c|}{------} \\
    &   & \multicolumn{2}{|c|}{[4 generations]}
    & \multicolumn{2}{|c|}{} \\
\cline{2-6}
 & 3 & $g^c$     & ($3^*$,1,1) & \multicolumn{2}{|c|}{------} \\
 &   & $l$       & (1,2,1)     & \multicolumn{2}{|c|}{------} \\
 &   & \multicolumn{2}{|c|}{[4 generations]} & \multicolumn{2}{|c|}{} \\
\cline{2-6}
 & 4 & $Q$       & (3,2,1)     & \multicolumn{2}{|c|}{------} \\
 &   & \multicolumn{2}{|c|}{[4 generations]} & \multicolumn{2}{|c|}{} \\
\hline \hline
 & 0 & $S$,$d^c$    & ($4^*$,1)
 & ${\overline S}$,${\overline d^c}$ & (4,1) \\
 &   & \multicolumn{2}{|c|}{[5 generations]} &
       \multicolumn{2}{|c|}{[1 anti-generation]} \\
\cline{2-6}
 & 1 & $H_d$ & (1,2) & \multicolumn{2}{|c|}{------} \\
 &   & $e^c$ & (1,1) & \multicolumn{2}{|c|}{------} \\
 &   & \multicolumn{2}{|c|}{[4 generations]}
 & \multicolumn{2}{|c|}{} \\
\cline{2-6}
(ii) & 2 & $u^c$,$g$     & (6,1)
& \multicolumn{2}{|c|}{------} \\
     &   & \multicolumn{2}{|c|}{[4 generations]}
     & \multicolumn{2}{|c|}{} \\
\cline{2-6}
 & 3 & $g^c$,$\nu ^c$     & ($4^*$,1)
 & \multicolumn{2}{|c|}{------} \\
 &   & $l$              & (1,2)
 & \multicolumn{2}{|c|}{------} \\
 &   & \multicolumn{2}{|c|}{[4 generations]}
 & \multicolumn{2}{|c|}{} \\
\cline{2-6}
 & 4 & $Q$,$H_u$          & (4,2)
 & \multicolumn{2}{|c|}{------} \\
 &   & \multicolumn{2}{|c|}{[4 generations]}
 & \multicolumn{2}{|c|}{} \\
\hline \hline
\end{tabular}

\newpage

{\bf Table 2} \\
\bigskip
\begin{tabular}{|c|c|c|c|} \hline
       fields        & $M_C - M_I$ & $M_I - M_S$ & $M_S - M_Z$ \\
\hline
       $d^c$                  &  5  &  5  &  4   \\
       $S$                    &  5  &  5  &  0   \\
       ${\overline d^c},{\overline S}$
                              &  1  &  1  &  0   \\
       $H_u,H_d$              &  4  &  1  &  1   \\
       $g,g^c$                &  4  &  0  &  0   \\
       $Q,u^c$                &  4  &  4  &  4   \\
       $l,\nu ^c,e^c$         &  4  &  4  &  3   \\
\hline
\end{tabular}

\vspace{4cm}

{\bf Table 3} \\
\bigskip
\begin{tabular}{|cc|ccc|cc|cc|} \hline
       fields    &    &  $2\sqrt {6}Q_{\psi }$
       &  $2\sqrt {10}Q_{\chi }$
                                    &  $2\sqrt {15}Q_{\eta }$
                   &  $2{\sqrt 6}Q_{\gamma }$
                   &  $2Q_{\delta }$
                    &  $2\sqrt {7}Q_{\theta }$
                    &  $2\sqrt {42}Q_{\xi }$  \\
\hline
       $(d^c,S)$       &  $(4^*,1)$
                         &   (1,4)  &   (3,0)   &   $(-1,5)$
                           &   1    &  $-1$ &  0   &  $-7$   \\
       $(g^c,\nu ^c)$  &  $(4^*,1)$
                         &  $(-2,1)$ &  $(-2,-5)$  &  $(-1,5)$
                           &   1    &   1   &  2   &   5     \\
       $(Q,H_u)$       &   (4,2)
                         &  $(1,-2)$  &  $(-1,2)$  &  $(2,-4)$
                           &   1    &   0   &  1   &  $-1$   \\
       $(u^c,g)$       &   (6,1)
                         &  $(1,-2)$  &  $(-1,2)$  &  $(2,-4)$
                           &  $-2$  &   0   & $-2$ &   2     \\
       $H_d$           &   (1,2)
                         &  $-2$   &   $-2$  &  $-1$
                           &  $-2$  &   1   & $-1$ &   8     \\
       $l$             &   (1,2)
                         &    1    &   3    &  $-1$
                           &  $-2$  &  $-1$ & $-3$ &  $-4$   \\
       $e^c$           &   (1,1)
                         &    1    &  $-1$  &   2
                           &   4    &   0   &  4   &  $-4$   \\
\hline
\end{tabular}

\newpage

\vspace {3cm}

{\bf Table 4} \\
\bigskip

\begin{tabular}{|c|cc|cc|c|} \hline
       fields      &  $2\sqrt {42}Q_{\xi }$  &  $2\sqrt {6}T_{15}$
                    &  $2\sqrt {42}Q_{\rho }$
                    &  $2\sqrt {7}Q_{\theta }$
                     &  $2\sqrt {15}Q_Y$   \\
\hline
       $d^c$        &  $-7$  &  $-1$  &  $-7$  &   0    &    2    \\
       $S$          &  $-7$  &   3    &   0    &   0    &    0    \\
       $g^c$        &   5    &  $-1$  &   2    &   2    &    2    \\
       $\nu ^c$     &   5    &   3    &   9    &   2    &    0    \\
       $Q$          &  $-1$  &   1    &   1    &   1    &    1    \\
       $H_u$        &  $-1$  &  $-3$  &  $-6$  &   1    &    3    \\
       $u^c$        &   2    &   2    &   5    &  $-2$  &  $-4$   \\
       $g$          &   2    &  $-2$  &  $-2$  &  $-2$  &  $-2$   \\
       $H_d$        &   8    &   0    &   6    &  $-1$  &  $-3$   \\
       $l$          &  $-4$  &   0    &  $-3$  &  $-3$  &  $-3$   \\
       $e^c$        &  $-4$  &   0    &  $-3$  &   4    &    6    \\
\hline
\end{tabular}

\newpage

\vspace {3cm}

{\bf Table 5} \\
\bigskip

\begin{tabular}{|cc||cccc|}\hline
\multicolumn{2}{|c||}{\raisebox{-1.9ex}{$M_C$}}
        & $U(1)_\gamma$ & $U(1)_\delta$ & $SU(4)_C$ & $SU(2)_L$ \\
      & & $U(1)_\theta$ & $U(1)_\xi $   & $|$ & $|$ \\
\cline{1-2}
 $\|$  &{\raisebox{-1.7ex}{$b_{ij}$}} & $|$ & $|$ & $|$ & $|$ \\
 $\Downarrow$   & & $|$ & $\downarrow$ & $\downarrow$ & $|$ \\
\cline{1-2}
      & & $\downarrow$ & $U(1)_\xi$ & $SU(4)_C$ & $|$\\
\multicolumn{2}{|c||}{$M_I$}
        & $U(1)_\theta$ & $U(1)_\rho$   & $SU(3)_C$ & $|$ \\
      & & $U(1)_\sigma$ & $U(1)_\tau $  & $|$ & $|$ \\
\cline{1-2}
  $\|$    &           & $|$ & $|$ & $|$ & $|$ \\
  $\|$    &{\raisebox{-1.7ex}{$b'_{ij}$}} & $|$ & $|$ & $|$ & $|$ \\
  $\|$    &           & $|$ & $|$ & $|$ & $|$ \\
  $\Downarrow$    &   & $\downarrow$ & $\downarrow$ & $|$ & $|$ \\
\cline{1-2}
\multicolumn{2}{|c||}{\raisebox{-1.9ex}{$M_S$}}
            & $U(1)_\sigma$   &    $U(1)_\tau$   & $|$ & $|$ \\
          & & \multicolumn{2}{c}{\quad $U(1)_Y$}    & $|$ & $|$ \\
\cline{1-2}
  $\Downarrow$   & $b''_{ij}$ &  \multicolumn{2}{c}{$\quad \downarrow$}
                                      & $\downarrow$ & $\downarrow$ \\
\cline{1-2}
\multicolumn{2}{|c||}{$M_Z$} &
     \multicolumn{2}{c}{\quad $U(1)_Y$}    & $SU(3)_C$ & $SU(2)_L$ \\
\hline
\end{tabular}

\end{center}


\newpage

\begin{thebibliography}{1}

\bibitem{Amaldi}
U.Amaldi,W.de Boer and H.Furstenau, Phys. Lett. {\bf 260B} (1991) 447. \\
P.Langacker and M.Luo, Phys. Rev. {\bf D44} (1991) 817. \\
J.Ellis, S.Kelly and D.V.Nanopoulos, Phys.Lett. {\bf 249B} (1990) 441.

\bibitem{Tsunoda}
T.Matsuoka, D.Mochinaga and K.Tsunoda, Nagoya Univ. preprint (1993)
DPNU-93-15 (to be published in Prog. Theor. Phys. {\bf 90} (1993) No.3).

\bibitem{4generation}
E.Witten, Nucl. Phys. {\bf B258} (1985) 75. \\
D.Gepner, Phys. Lett. {\bf 199B} (1987) 370; Nucl. Phys. {\bf B296}
 (1988) 757;  {\bf B311} (1988) 191. \\
M.Matsuda, T.Matsuoka, H.Mino,
D.Suematsu and Y.Yamada, Prog. Theor. Phys.
{\bf 79} (1988) 174. \\
D.Suematsu, Phys. Rev. {\bf D38} (1988) 3128. \\
P.Zoglin, Phys. Lett. {\bf 228B} (1989) 47.

\bibitem{Choi}
K.Choi and J.E.Kim, Phys. Lett. {\bf 165B} (1985) 71. \\
T.Matsuoka and D.Suematsu, Prig. Theor. Phys. {\bf 76} (1986) 901.

\bibitem{Kikkawa}
K.Kikkawa and M.Yamasaki, Phys. Lett. {\bf 114B} (1984) 357. \\
N.Sakai and I.Senda, Prog. Theor. Phys. {\bf 75} (1986) 692. \\
V.P.Nair, A.Shapere, A.Strominger and F.Wilczek, Nucl. Phys. {\bf B287}
           (1987) 402 \\
B. Sathiapalan, Phys. Rev. Lett. {\bf 58} (1987).

\bibitem{Kaplunovsky}
V.S.Kaplunovsky, Nucl. Phys. {\bf B307} (1988) 145. \\
L.J.Dixon, V.S.Kaplunovsky and J.Louis.
Nucl. Phys. {\bf B355} (1991) 649. \\
I.Antoniadis, K.S.Narain and T.R.Taylor,
Phys. Lett. {\bf 267B} (1991) 37.

\bibitem{Kalara}
S.Kalara, J.L.Lopez and D.V.Nanopoulos,
Phys. Lett. {\bf 287B} (1992) 82.

\bibitem{Matsuoka}
T.Matsuoka and D.Suematsu, Nucl. Phys. {\bf B274} (1986) 106;
Prog. Theor. Phys. {\bf 76} (1986) 886.

\bibitem{Jones}
D.R.T.Jones, Phys. Rev. {\bf D25} (1982) 581. \\
The last term of {\it r.h.s.} in Eq.(35),
which is not considered in this reference,
is brought about
by the effect of $U(1)$ kinetic mixings.
This is the characteristic feature in the case of the models
with more than one $U(1)$ gauge group.

\bibitem{CHARGINO}
C.Hattori, M.Matsuda, T.Matsuoka and D.Mochinaga, Prog. Theor. Phys.
{\bf 86} (1991) 725.

\bibitem{PDG}
K.Hikasa et al. [Particle Data Group], Phys. Rev. {\bf D45} (1992) S1.

\bibitem{3gener}
B.R.Greene, K.H.Kirklin, P.J.Miron and G.G.Ross,
Phys. Lett. {\bf 180B}
(1986) 69; Nucl. Phys. {\bf B278} (1986) 667;
{\bf B292} (1987) 606.

\bibitem{Dela}
F.Delaguila, G.D.Coughlan and M.Masip,
Phys. Lett. {\bf 227B} (1989) 55.

\end{thebibliography}
\end{document}